\documentclass[runningheads]{llncs}
\usepackage[T1]{fontenc}

\usepackage{graphicx}
\usepackage{amsmath,amssymb,mathtools}
\usepackage{dsfont}
\usepackage{booktabs}  
\usepackage{siunitx}
\usepackage{hyperref}
\usepackage{subcaption}
\usepackage{multirow}
\usepackage{yfonts}

\usepackage{color}

\usepackage{xcolor}

\begin{document}
\title{When Attention Becomes Exposure \\ in  Generative Search \vspace{-0.3in}}

\author{Shayan Alipour\inst{1,2} \and
Mehdi Kargar\inst{1} \and
Morteza Zihayat\inst{1}}

\institute{Toronto Metropolitan University
\email{\{shayan.alipour,kargar,mzihayat\}@torontomu.ca } \and
Sapienza University of Rome
}

\maketitle  

\begin{abstract}
Generative search engines are reshaping information access by replacing traditional ranked lists with synthesized answers and references. In parallel, with the growth of Web3 platforms, incentive-driven creator ecosystems have become an essential part of how enterprises build visibility and community by rewarding creators for contributing to shared narratives. However, the extent to which exposure in generative search engine citations is shaped by external attention markets remains uncertain. In this study, we audit the exposure for 44 Web3 enterprises. First, we show that the creator community around each enterprise is persistent over time. Second, enterprise-specific queries reveal that more popular voices systematically receive greater citation exposure than others. Third, we find that larger follower bases and enterprises with more concentrated creator cores are associated with higher-ranked exposure. Together, these results show that generative search engine citations exhibit exposure bias toward already prominent voices, which risks entrenching incumbents and narrowing viewpoint diversity. 

\vspace{-0.1in}
\end{abstract}

\section{Introduction}
\label{sec:introduction}
Generative Search Engines (GSEs) are reshaping information access from ranked lists of links toward synthesized answers with citation panels~\cite{allan2024future,kaiser2025new,suri2024use}. Unlike traditional Search Engine Results Pages (SERPs), where exposure depends on position, visibility in GSEs depends on which references are cited and how frequently they appear across responses~\cite{gienapp2024evaluating}. Understanding who is cited, and how exposure is distributed among sources, is therefore central to assessing credibility and fairness in this new retrieval paradigm~\cite{chen2025generative,aggarwal2024geo}. Previous studies have addressed selective exposure in IR systems extensively by formalizing diversification to counter concentration~\cite{carbonell1998use,agrawal2009diversifying,santos2010explicit} and developing frameworks to reason about exposure as a resource that can be made more equitable across items and groups~\cite{singh2018fairness,biega2018equity,diaz2020evaluating,helberger2012exposure}. Building on this foundation, early audits suggest that GSEs exhibit agreement-seeking tendencies and may narrow exposure toward favored viewpoints~\cite{sharma2024generative,venkit2024search}. In parallel, the broader digital landscape has evolved into an attention economy, where visibility itself operates as a form of currency. This economy has intensified through content-creator platforms and algorithmic ranking systems that reward engagement~\cite{kristoufek2013bitcoin,li2024impact}. Communities within Web3---broadly defined as decentralized ecosystems built on blockchain protocols where ownership and participation are often tokenized---represent a particularly organized and measurable example. Within this landscape, organizations developing decentralized protocols or applications, referred to as \emph{enterprises} in this study, rely on networks of incentivized creators who promote updates and narratives. Recent studies discuss new strategies that the creator economy can adopt to increase exposure~\cite{aggarwal2024geo,wu2025generative}. These trends raise concerns about exposure bias, in which GSEs surface the same popular voices, thereby reinforcing their standing. 

In this study, we audit how the attention economy shapes diversity and exposure in generative search. We measure \textbf{RQ1}: how concentrated Web3's creator ecosystem is, \textbf{RQ2:} to what extent do GSEs cite these creators, and if external prominence increases citation exposure, and \textbf{RQ3:} which factors explain higher citation visibility. Our findings indicate that creator communities are stable and hierarchical, prominent creators receive significantly higher citation exposure, and exposure correlates with follower counts and internal concentration. We further observe that GSEs' citation panels differ by access method, with API calls consistently returning fewer and more limited panels than the user interface (UI). Together, these findings show that GSEs amplify existing hierarchies of attention, raising implications for viewpoint diversity, equitable information access, and the design of responsible generative retrieval systems. We contribute a curated corpus of queries, creator leaderboards, GSE responses with citation panels, and supporting analysis materials for further community use.\footnote{\url{https://github.com/shayanalipour/attention_becomes_exposure}
}

\section{Methodology}
\label{sec:methods}

\textbf{a) Data and Experimental Context.} We analyze $44$ Web3 enterprises representing major protocols, infrastructure providers, and decentralized applications. Each enterprise maintains an active creator ecosystem on X (formerly Twitter), where creators promote updates, discussions, and narratives surrounding decentralized technologies. To measure creator prominence, we use publicly available leaderboards from \emph{KaitoAI}\footnote{\url{https://yaps.kaito.ai}}, which rank contributors by a proprietary \emph{Mindshare} score capturing content quality, engagement authenticity, and consistency within each enterprise community. We collect the Top–100 ranked creators per enterprise from two leaderboard snapshots taken within a three–week interval (2025-09-28 and 2025-10-18). These snapshots enable the assessment of temporal persistence and rank stability in creator communities (Unique creators=$3{,}232$). To evaluate visibility in GSEs, we construct three queries per enterprise (two general, one time-sensitive) which yields $132$ unique queries. General queries target the enterprise’s purpose or core technology, while time-sensitive ones reference recent announcements within the prior 10 days sourced from official channels.

\noindent\textbf{b) Audit Design.} We use three enterprise-specific queries and examine four GSEs, consisting of Grok-4 with Live Search (Grok), GPT-5 with web search (GPT), Gemini 2.5 Flash with Google grounding (Gemini), and Perplexity Sonar-Pro with web search (Perplexity). During pilot runs for GPT, Gemini, and Perplexity, we observed discrepancies between API and UI outputs, in which the API calls returned limited citation panels (community issues recognized\footnote{\url{https://x.com/nikunjhanda/status/1912387834234958203}}). To quantify the impact, we conduct a separate audit for Perplexity by comparing UI and API citation panels for identical queries. Finally, to assess creators’ exposure in isolation, we examine another setup that restricts Grok’s sources to content from the X platform.

\noindent\textbf{c) Exposure Quantification Measures.} Following head/tail analyses in recommender systems~\cite{abdollahpouri2019managing}, we compare exposure for top versus lower-ranked creators at the enterprise level, where for an enterprise \(e\) with \(N_e\) ranked creators (lower rank is better), define the exposure advantage of the Top-10 relative to the bottom half as 
\[
\Delta_e = \frac{1}{10}\sum_{r=1}^{10}\text{share}_{e,r} \;-\; \frac{1}{N_e-50}\sum_{r=51}^{N_e}\text{share}_{e,r},
\]
where \(\text{share}_{e,r}\) is the fraction of all citations in enterprise \(e\) attributed to the creator at rank \(r\). 
Positive \(\Delta_e\) indicates that Top-10 creators receive a larger per-rank share than the bottom half. 
We summarize across enterprises by estimating the mean \(\mathbb{E}[\Delta_e]\) and its 95\% confidence interval via bootstrap resampling over enterprises (\(B=1{,}000\)).

To measure deviation from a random baseline, we use the position-agnostic Normalized Cumulative Gain (NCG)~\cite{jarvelin2002cumulated}, which assigns graded credit by external rank and compares observed citation panels to an enterprise-specific expectation. For an enterprise $e$, each query $q$ yields an unordered citation panel $S_{e,q}$ of size $m_{e,q}=|S_{e,q}|$. We define the graded gain $g(r)$ in \eqref{eq:gain}. The panel's observed cumulative gain is \eqref{eq:gobs} where $r_e(i)$ is creator $i$'s external rank within enterprise $e$. We then normalize by the extreme case where all cited items are Top-10 creators \eqref{eq:gstar}, which upper-bounds any citation panel. Given that, we calculate the NCG in \eqref{eq:ncg}.

\begin{minipage}[t]{0.48\linewidth}
\begin{equation}
g(r)=
\begin{cases}
3,& r\le 10\\[2pt]
2,& 11\le r\le 50\\[2pt]
1,& 51\le r\le 100\\[2pt]
0,& r>100
\end{cases}
\label{eq:gain}
\tag{1}
\end{equation}
\end{minipage}\hfill
\begin{minipage}[t]{0.48\linewidth}
\begin{align}
G^{\mathrm{obs}}_{e,q}
&= \sum_{i\in S_{e,q}} g\!\big(r_e(i)\big),
\tag{2}\label{eq:gobs}\\
G^{\star}_{e,q}
&= 3\,m_{e,q},
\tag{3}\label{eq:gstar}
\end{align}
\end{minipage}

\begin{equation}
\mathrm{NCG}^{\mathrm{obs}}_{e,q}
= \frac{G^{\mathrm{obs}}_{e,q}}{G^{\star}_{e,q}}
= \frac{\sum_{i\in S_{e,q}} g\!\big(r_e(i)\big)}{3\,m_{e,q}}
\in[0,1].
\tag{4}\label{eq:ncg}
\end{equation}

To quantify amplification beyond chance, we compare $\mathrm{NCG}^{\text{obs}}$ to the expected value under citation panels drawn uniformly at random from the enterprise's candidate set of size $N_e$. Let $\bar g_e = \frac{1}{N_e} \sum_{r=1}^{N_e} g(r)$ and $g_{\max}=3$. Since the expected sum of gains over $m_{e,q}$ draws is $m_{e,q}\,\bar g_e$, the expected NCG is $\frac{\bar g_e}{g_{\max}}$. For each $(e,q)$ we compute the uplift $\Delta_{e,q} = \mathrm{NCG}^{\text{obs}}_{e,q} - \bar g_e/g_{\max}$. We then average across queries for that enterprise to obtain an enterprise-level effect $\Delta_e$, and summarize across enterprises by the mean of $\{\Delta_e\}$ with 95\% CIs via bootstrap resampling over enterprises ($B=1{,}000$). We report the mean uplift, its intervals, and percentage points (pp), where $\Delta_{\text{pp}}=100\,\Delta$.

\noindent\textbf{d) Modeling Exposure Associations.} To examine which factors are associated with higher exposure, we test whether (i) the concentration of participation within an enterprise, measured by Jaccard similarity $j$, and (ii) a creator’s audience size $f$ (followers), are associated with their rank. We estimate OLS with robust clustered standard errors:
\[
\log(r_i) \;=\; \beta_0 \;+\; \beta_1\, j_i \;+\; \beta_2\, \log(1+f_i) \;+\; \varepsilon_i,
\]
where the dependent variable is the log of the creator’s rank (lower $r$ is better), so negative coefficients indicate higher-ranked exposure. The sample is restricted to citation-panel sources that match ranked creators in enterprise leaderboards. Predictors are mean-centered so the intercept is evaluated at average $j$ and $f$.

\section{Findings}
\label{sec:findings}

\begin{figure}[t]
    \centering
    \begin{subfigure}{0.45\linewidth}
        \centering
        \caption{Temporal persistency of creators}
        \includegraphics[width=\linewidth]
        {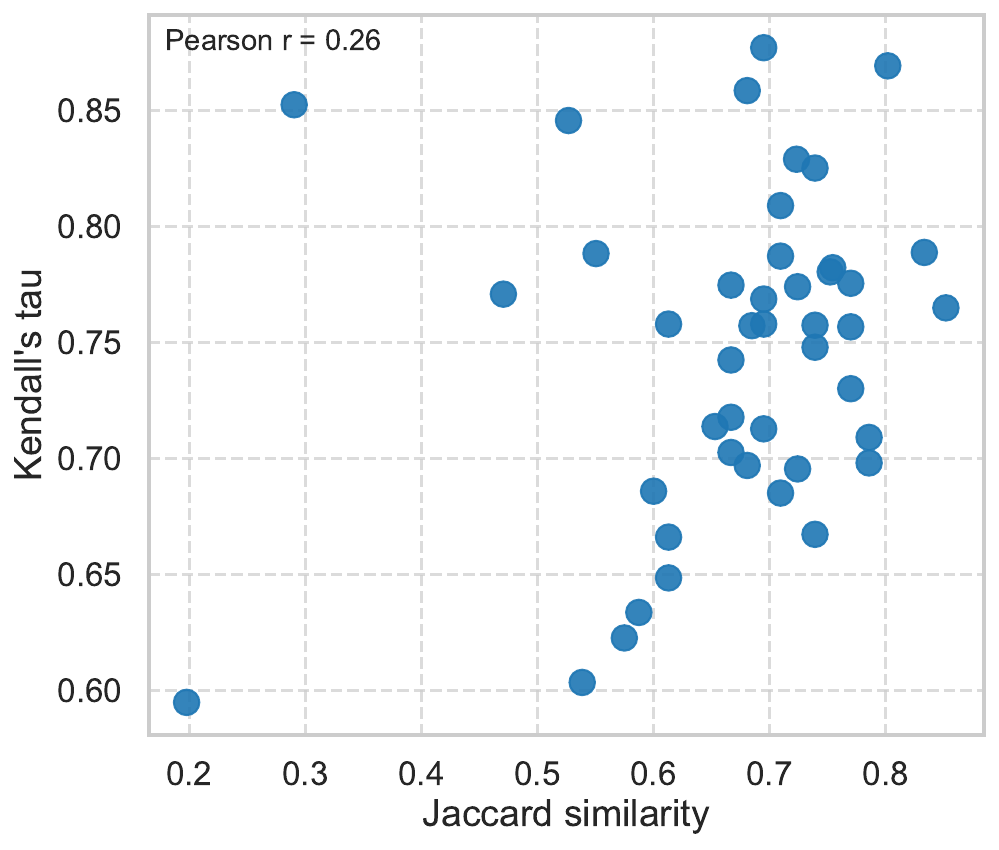}
        \label{fig:yapper_diversity}\vspace{-0.1in}
    \end{subfigure}
    \hfill
    \begin{subfigure}{0.45\linewidth}
        \centering
        \caption{Distribution of cited creators' ranks}
        \includegraphics[width=\linewidth]{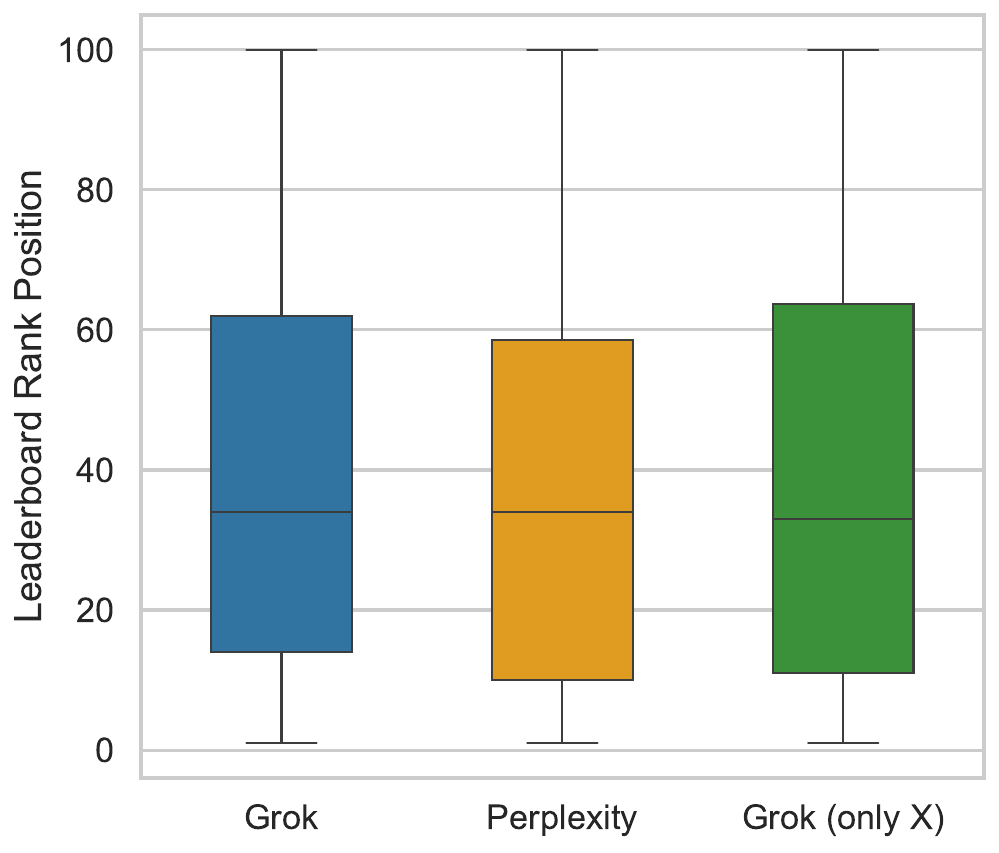}
        \label{fig:rank_dist}\vspace{-0.1in}
    \end{subfigure}
    \caption{(a) Membership and rank-order stability of creator communities across enterprises over a 20-day interval. (b) Distribution of cited creator ranks.}
    \label{fig:diversity_n_cite_dist}\vspace{-2em}
\end{figure}

\textbf{RQ1: Diversity of Creators.}
For each enterprise, we compute the membership stability of the Top-100 set with the Jaccard similarity $J$, where $J\in[0,1]$ and where 0 indicates no shared creators and 1 indicates identical leaderboards. We asses rank-order stability within that set using Kendall's $\tau$ applied to the intersection of creators observed in both snapshots, where $\tau\in[-1,1]$ and where 1 indicates perfect agreement, 0 indicates no association, and $-1$ indicates complete reversal. Over this 20-day interval, creator communities exhibit high stability in both membership and ranks, indicating that the same core of creators remain active and their relative influence is largely preserved over the period. Figure~\ref{fig:yapper_diversity} summarizes these distributions and their joint relationship where on average across enterprises, the Jaccard similarity equals 0.67 and Kendall's $\tau$ equals 0.75 for all creators.

\vspace{-1.5em}
\begin{table}[!htbp]
    \centering
    \sffamily
    \caption{Breakdown of citation panels across enterprises per GSEs}
    \begin{tabular}{lrrrr}
        \toprule
        \multicolumn{1}{c}{} & Queries & Unique URLs & X URLs (n, \%) & Handles (n, \%) \\
        \midrule
        Grok            & 132 & 2,223 & 1,123(50.52\%) & 152 (7.02\%) \\
        Perplexity (UI) & 132 & 3,698 & 1,255(33.94\%) & 134 (3.62\%) \\
        Grok (only X)   & 132 & 2,194 & 2,194(100\%)   & 333 (15.12\%) \\
        \bottomrule
    \end{tabular}
    \label{tab:inference}
\end{table}

\vspace{-1em}

\textbf{RQ2: Creator Exposure.} 
We next analyze how GSEs allocate exposure to creators across enterprises. Each query–GSE pair returns a citation panel, which we parse and match to ranked creators. We begin by contrasting retrieval configurations, where GSEs with direct access to social sources (e.g., Grok’s “Web + X” and “X-only” modes) exhibit greater coverage of creator-linked citations than systems grounded primarily in web content. In contrast, API configurations for GPT, Gemini, and Perplexity return no sources from X, even when the domain is explicitly requested in the parameters. However, in Perplexity’s UI, a baseline run using a single query per enterprise (without domain-inclusion controls) showed $12.13\%$ X coverage. To align Perplexity more closely with Grok’s source mix, we ran three queries per enterprise and enabled domain inclusion for X alongside web access. Under this setting, Perplexity returned $3{,}698$ unique sources, of which $1{,}255$ were from X ($33.94\%$) with $134$ matched creator handles ($3.62\%$). Finally, restricting Grok’s sources to X only yields $2{,}194$ X citations, with $333$ overlapping creator handles ($15.12\%$). Table~\ref{tab:inference} reports the citation-panel breakdown per GSE.

To quantify exposure concentration, we compute within-enterprise contrasts between Top-10 and tail creators where positive values indicate an advantage for higher-ranked creators. Grok shows a cross-enterprise mean difference of $+4.32$ percentage points (pp) with a $95\%$ bootstrap interval $[+1.46,\,+7.51]$. Perplexity exhibits a similar pattern with a mean advantage of $+4.79$~pp $[+2.42,\,+7.35]$. When Grok is limited to X-only sources, the advantage increases to $+7.16$~pp, which indicates stronger concentration among the most prominent voices. These estimates are summarized in Table~\ref{tab:exposure_metrics} and are consistent with the distributions of cited ranks in Figure~\ref{fig:rank_dist}, which are skewed toward higher-ranked (lower-numbered) creators across GSEs.

To assess amplification beyond chance, we compare the observed NCG to an enterprise-specific random baseline equal to the expected normalized gain under uniform draws from the enterprise's ranked creator set. We report uplift as the difference (observed minus baseline) in percentage points. Across configurations, we observe significant uplift beyond chance. For Grok, average NCG uplift is $+9.39$~pp with a $95\%$ interval $[+2.59,\,+16.45]$ across $n=37$ enterprises (excluded enterprises are due to missing overlap). Perplexity and Grok (X-only) show comparable uplifts of $+9.94$~pp and $+8.99$~pp, respectively (See Table~\ref{tab:exposure_metrics})

\textbf{RQ3: Drivers of Exposure.}
We examine associations between creator and enterprise attributes and citation visibility. Table~\ref{tab:exposure_drivers} reports coefficients from an OLS regression of citation rank on enterprise-level concentration and creator audience size. The intercept ($\beta_0=3.05$) provides a baseline corresponding to an expected rank of about 21 at average levels of follower and diversity metrics. Relative to this baseline, results indicate that exposure is biased toward creators with larger audiences and enterprises whose creator activity is more concentrated. In particular, follower counts have a significant association ($\beta_2=-0.44$) on the exposure of higher-ranked creators in citation panels. Enterprises' creator community diversity, measured by Jaccard similarity $j$ ($\beta_1=-1.62$), which reflects on membership stability, suggests that more concentrated creator ecosystems increase the likelihood of higher-ranked voices receiving coverage.

\vspace{-1em}

\begin{table}[!htbp]
    \centering
    \sffamily\footnotesize
    \caption{(a) Top-10 advantage and NCG uplift across GSEs, reported as mean percentage-point differences. (b) OLS coefficients for citation exposure.}
    \begin{subtable}[t]{0.6\linewidth}
        \begin{tabular}{rccc}
        \toprule
         & Top-10 vs tail ($95\%$ CI) & NCG uplift ($95\%$ CI) & Enterprises \\
        \midrule
        Grok & 4.32 (1.46,7.51) & 9.39 (2.59,16.45) & 37 \\
        \midrule
        Perplexity & 4.79 (2.42,7.35) & 9.94 (3.18,16.57) & 38 \\
        \midrule
        Grok(X) & 7.16 (3.81,10.93) & 8.99 (3.83,14.28) & 41\\
        \bottomrule \\[-1.8ex]
    \end{tabular}
    \caption{}
    \label{tab:exposure_metrics}
    \end{subtable}
    \hfill
    \begin{subtable}[t]{0.2\linewidth}
        \begin{tabular}{lc}
        \toprule 
        & Coef. \\
        \midrule
        Intercept & 3.05$^{*}$ \\
        $j_i$  & -1.62$^{*}$ \\
        $\log(1+f_i)$ & -0.44$^{*}$ \\
        \midrule
        observations & 373 \\
        pseudo $R^2$ & 0.30 \\
        \midrule
        \multicolumn{2}{l}{$^{*}$\; $p<0.05$} \\
        \bottomrule \\[-1.8ex]
    \end{tabular}
    \caption{}
    \label{tab:exposure_drivers}
    \end{subtable}
    \label{tab:rq2_rq3}\vspace{-4em}
\end{table}

\vspace{-1em}

\section{Implications and Theory of Change}
\label{sec:implications}

Generative search engines replace ranked lists with synthesized answers and citations that shift how visibility is allocated. In creator-driven ecosystems like Web3, where attention already concentrates around a few voices, this can amplify existing hierarchies and reduce viewpoint diversity. We address these concerns by contributing (i) an audit framework that treats citation panels as exposure allocations, (ii) evidence of coverage gaps between API and UI outputs, (iii) enterprise-level estimates showing a Top-10 advantage and amplification beyond chance, and (iv) a corpus comprising enterprise-specific queries, creator leaderboards, and citation panels. Together, these contributions enable more rigorous assessment of exposure bias in future studies. We call on different stakeholder groups---including GSE companies and researchers---to increase efforts toward API--UI parity and, in the meantime, to account for these differences when designing audits, reporting results, and comparing findings across studies.

\textbf{Limitations.}
Our audit focuses on Web3, which may limit generalizability to other domains where attention dynamics differ~\cite{di2025patterns,alipour2024cross}. While we document the API--UI gap, we cannot fully attribute its causes due to limited visibility into system internals. Creator prominence is measured with an external leaderboard whose scoring rubric may embed its own biases. Finally, our exposure metrics assume uniform attention across citations and do not incorporate interaction logs, so they estimate potential exposure rather than realized impact. 

\bibliographystyle{splncs04}
\bibliography{main}

@inproceedings{gienapp2024evaluating,
  title={Evaluating generative ad hoc information retrieval},
  author={Gienapp, Lukas and Scells, Harrisen and Deckers, Niklas and Bevendorff, Janek and Wang, Shuai and Kiesel, Johannes and Syed, Shahbaz and Fr{\"o}be, Maik and Zuccon, Guido and Stein, Benno and others},
  booktitle={Proceedings of the 47th International ACM SIGIR Conference on Research and Development in Information Retrieval},
  pages={1916--1929},
  year={2024}
}

@article{allan2024future,
  title={Future of Information Retrieval Research in the Age of Generative AI},
  author={Allan, James and Choi, Eunsol and Lopresti, Daniel P and Zamani, Hamed},
  journal={arXiv preprint arXiv:2412.02043},
  year={2024}
}

@article{suri2024use,
  title={The use of generative search engines for knowledge work and complex tasks},
  author={Suri, Siddharth and Counts, Scott and Wang, Leijie and Chen, Chacha and Wan, Mengting and Safavi, Tara and Neville, Jennifer and Shah, Chirag and White, Ryen W and Andersen, Reid and others},
  journal={arXiv preprint arXiv:2404.04268},
  year={2024}
}

@inproceedings{carbonell1998use,
  title={The use of MMR, diversity-based reranking for reordering documents and producing summaries},
  author={Carbonell, Jaime and Goldstein, Jade},
  booktitle={Proceedings of the 21st annual international ACM SIGIR conference on Research and development in information retrieval},
  pages={335--336},
  year={1998}
}

@inproceedings{agrawal2009diversifying,
  title={Diversifying search results},
  author={Agrawal, Rakesh and Gollapudi, Sreenivas and Halverson, Alan and Ieong, Samuel},
  booktitle={Proceedings of the second ACM international conference on web search and data mining},
  pages={5--14},
  year={2009}
}

@inproceedings{santos2010explicit,
  title={Explicit search result diversification through sub-queries},
  author={Santos, Rodrygo LT and Peng, Jie and Macdonald, Craig and Ounis, Iadh},
  booktitle={European conference on information retrieval},
  pages={87--99},
  year={2010},
  organization={Springer}
}

@inproceedings{singh2018fairness,
  title={Fairness of exposure in rankings},
  author={Singh, Ashudeep and Joachims, Thorsten},
  booktitle={Proceedings of the 24th ACM SIGKDD international conference on knowledge discovery \& data mining},
  pages={2219--2228},
  year={2018}
}

@inproceedings{biega2018equity,
  title={Equity of attention: Amortizing individual fairness in rankings},
  author={Biega, Asia J and Gummadi, Krishna P and Weikum, Gerhard},
  booktitle={The 41st international acm sigir conference on research \& development in information retrieval},
  pages={405--414},
  year={2018}
}

@inproceedings{diaz2020evaluating,
  title={Evaluating stochastic rankings with expected exposure},
  author={Diaz, Fernando and Mitra, Bhaskar and Ekstrand, Michael D and Biega, Asia J and Carterette, Ben},
  booktitle={Proceedings of the 29th ACM international conference on information \& knowledge management},
  pages={275--284},
  year={2020}
}

@article{helberger2012exposure,
  title={Exposure diversity as a policy goal},
  author={Helberger, Natali},
  journal={Journal of Media Law},
  volume={4},
  number={1},
  pages={65--92},
  year={2012},
  publisher={Taylor \& Francis}
}

@inproceedings{sharma2024generative,
  title={Generative echo chamber? effect of llm-powered search systems on diverse information seeking},
  author={Sharma, Nikhil and Liao, Q Vera and Xiao, Ziang},
  booktitle={Proceedings of the 2024 CHI Conference on Human Factors in Computing Systems},
  pages={1--17},
  year={2024}
}

@article{venkit2024search,
  title={Search engines in an ai era: The false promise of factual and verifiable source-cited responses},
  author={Venkit, Pranav Narayanan and Laban, Philippe and Zhou, Yilun and Mao, Yixin and Wu, Chien-Sheng},
  journal={arXiv preprint arXiv:2410.22349},
  year={2024}
}

@article{abdollahpouri2019managing,
  title={Managing popularity bias in recommender systems with personalized re-ranking},
  author={Abdollahpouri, Himan and Burke, Robin and Mobasher, Bamshad},
  journal={arXiv preprint arXiv:1901.07555},
  year={2019}
}

@article{chen2025generative,
  title={Generative engine optimization: How to dominate ai search},
  author={Chen, Mahe and Wang, Xiaoxuan and Chen, Kaiwen and Koudas, Nick},
  journal={arXiv preprint arXiv:2509.08919},
  year={2025}
}

@inproceedings{aggarwal2024geo,
  title={Geo: Generative engine optimization},
  author={Aggarwal, Pranjal and Murahari, Vishvak and Rajpurohit, Tanmay and Kalyan, Ashwin and Narasimhan, Karthik and Deshpande, Ameet},
  booktitle={Proceedings of the 30th ACM SIGKDD Conference on Knowledge Discovery and Data Mining},
  pages={5--16},
  year={2024}
}

@inproceedings{kaiser2025new,
  title={A New Era of Online Search? A Large-Scale Study of User Behavior and Personal Preferences during Practical Search Tasks with Generative AI versus Traditional Search Engines},
  author={Kaiser, Carolin and Kaiser, Jakob and Schallner, Rene and Schneider, Sabrina},
  booktitle={Proceedings of the Extended Abstracts of the CHI Conference on Human Factors in Computing Systems},
  pages={1--7},
  year={2025}
}

@article{wu2025generative,
  title={What Generative Search Engines Like and How to Optimize Web Content Cooperatively},
  author={Wu, Yujiang and Zhong, Shanshan and Kim, Yubin and Xiong, Chenyan},
  journal={arXiv preprint arXiv:2510.11438},
  year={2025}
}

@article{kristoufek2013bitcoin,
  title={BitCoin meets Google Trends and Wikipedia: Quantifying the relationship between phenomena of the Internet era},
  author={Kristoufek, Ladislav},
  journal={Scientific reports},
  volume={3},
  number={1},
  pages={3415},
  year={2013},
  publisher={Nature Publishing Group UK London}
}

@article{di2025patterns,
  title={Patterns, Models, and Challenges in Online Social Media: A Survey},
  author={Di Marco, Niccol{\`o} and Bonetti, Anita and Di Martino, Edoardo and Loru, Edoardo and Nudo, Jacopo and Pandolfo, Mario Edoardo and Pecile, Giulio and Sangiorgio, Emanuele and Scalco, Irene and Zollo, Simon and others},
  journal={arXiv preprint arXiv:2507.13379},
  year={2025}
}

@article{alipour2024cross,
  title={Cross-platform social dynamics: an analysis of ChatGPT and COVID-19 vaccine conversations},
  author={Alipour, Shayan and Galeazzi, Alessandro and Sangiorgio, Emanuele and Avalle, Michele and Bojic, Ljubisa and Cinelli, Matteo and Quattrociocchi, Walter},
  journal={Scientific Reports},
  volume={14},
  number={1},
  pages={2789},
  year={2024},
  publisher={Nature Publishing Group UK London}
}

@article{li2024impact,
  title={The impact of sentiment and engagement of Twitter posts on cryptocurrency price movement},
  author={Li, Scott and Ma, Judy},
  journal={Finance Research Letters},
  volume={65},
  pages={105598},
  year={2024},
  publisher={Elsevier}
}

@article{jarvelin2002cumulated,
  title={Cumulated gain-based evaluation of IR techniques},
  author={J{\"a}rvelin, Kalervo and Kek{\"a}l{\"a}inen, Jaana},
  journal={ACM Transactions on Information Systems (TOIS)},
  volume={20},
  number={4},
  pages={422--446},
  year={2002},
  publisher={ACM New York, NY, USA}
}
\end{document}